\documentstyle[11pt,newpasp,twoside,epsf,graphics]{article}
\def\edcomment#1{\iffalse\marginpar{\raggedright\sl#1\/}\else\relax\fi}
\reversemarginpar

\begin{document}

\title{Intermediate Element Abundances in Galaxy Clusters}

\author{W. H. Baumgartner\footnotemark}\footnotetext{Email: {\tt wayne@astro.umd.edu}}
\author{R. F. Mushotzky}
\affil{Laboratory for High Energy Astrophysics, NASA/GSFC, Code 662,
Greenbelt, MD 20771}

\author{D. J. Horner}
\affil{Space Telescope Science Institute, Baltimore, MD, 21218}

\begin{abstract}
\begin{itemize}
\item We have determined the elemental abundances of Fe, Si, S, Ar,
Ca, Ne, Mg, and Ni in the intra-cluster medium (ICM) using all the
clusters in the archives of the \textsl{ASCA} X-ray telescope.
\item The calcium and argon abundances are very low and are not
consistent with the abundances of the two well determined $\alpha$
elements, silicon and sulfur.
\item The results do not show a clear preference for metal
enrichment by solely Type Ia supernovae or Type II.  
\item Trends in the abundances as a function of temperature (mass)
suggest that different processes for enrichment and distribution of
metals are important on different size scales.
\end{itemize}
\end{abstract}

\section{Introduction}
The intra-cluster medium of galaxy clusters is the repository of all
the metals produced by the stars in member galaxies.  The
determination of the elemental abundances in clusters provides an
integrated measurement of metal production throughout the history of
the cluster.

The measurement of elemental abundances in clusters is in many ways
more straightforward than in other objects.  Clusters are optically
thin to X-rays and the spectra do not suffer from the complicating effects
of radiative transport and dust common in photospheric measurements and
galactic H\,II regions.

Compilations of X-ray cluster abundances exist (White 2000), but only
for the iron-dominated overall metal abundance in clusters.  Our
compilation is the first large catalog of intermediate element
abundances in clusters able to differentiate between several of the
$\alpha$ elements (Si, S, Ar, Ca), and the iron peak elements (Fe, Ni)
observable by X-ray instruments.  The results from this analysis are
average elemental abundances from ensembles of clusters with similar
properties.  They provide a more general view in contrast with
detailed spatially resolved abundance measurements for individual
clusters from \textsl{Chandra} and \textsl{XMM}.

\begin{figure}[t]
\resizebox{\textwidth}{!}{\rotatebox{90}{\includegraphics{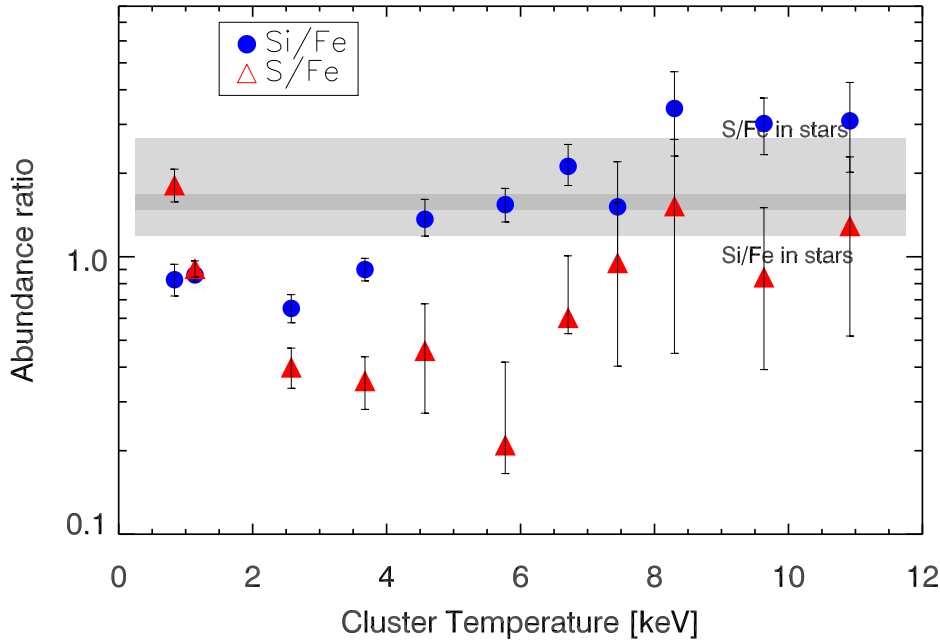}}}
\caption{This plot compares our [Si/Fe] and [S/Fe] ratios to those measured in
stars.  Stellar data is from Timmes, Woosley \& Weaver (1995); the
upper gray bar is stellar [S/Fe] data, and the lower gray bar is
stellar [Si/Fe] data.}
\label{starscompare}
\end{figure}

\section{Data Sample and Reduction}

We use the cluster sample of Horner et al.\ (ApJS submitted).  All of
the cluster observations ($\sim 300$) in the archives of the \textsl{ASCA}
X-ray satellite were homogeneously reduced and fit to a MEKAL
plasma model with galactic absorption to obtain the cluster
temperature and metal abundance. 

Individual cluster observations with \textsl{ASCA} do not have enough
signal to noise to allow a determination of the intermediate cluster
abundances.  In order to improve the signal and determine the
individual element abundances, we stacked together
single cluster observations with similar properties. The individual
cluster observations were divided into 22 bins based on their
temperature and iron abundance: There are eleven 1~keV wide bins from 0
to 11 keV, and each bin is equally split into a low Fe abundance group
and a high Fe abundance group.  There are 218 clusters in our sample.
We exclude clusters with more than 40 thousand counts in order not to
unduly bias the results in a single bin.  All of the clusters in a bin
(typically 8 -- 20) are jointly refit to a single VMEKAL model with
the same set of abundances for all the clusters in each bin. The solar
abundances of the elements we use are given in Grevesse \& Sauval
(1998).  The results from the high Fe bin and the low Fe bin are
combined to give a single data point for each 1~keV temperature bin.
We fit for Fe, S, Si, Ca, Ar, and Ni separately, and for Ne and Mg
tied together.

In order to assess the effects that a possible systematic error in the
\textsl{ASCA} calibration might have on our abundance determinations,
we carefully examined the residuals from a very long observation of
the quasar 3c273.  3c273 and sources in its class are known for 
featureless powerlaw spectra, allowing us to assume that any features
left after fitting can be attributed to calibration errors in the
response matrix.  We reduced the \textsl{ASCA} 3c273 data the same way
the cluster data was reduced, and fit the quasar with an absorbed
power law model.  We then took the residuals from the fit and applied
them to the cluster data in order to correct for any response matrix
errors, a sort of spectral flat-fielding.  We then refit the clusters
with the corrected spectra.  The resulting abundances are identical to
the original abundances within the statistical errors, showing that
response matrix problems cannot be the source of significant
systematic errors in the abundance determinations.

\section{Results}

\begin{figure}
\begin{center}
\resizebox{0.975\textwidth}{!}{\rotatebox{0}{\includegraphics{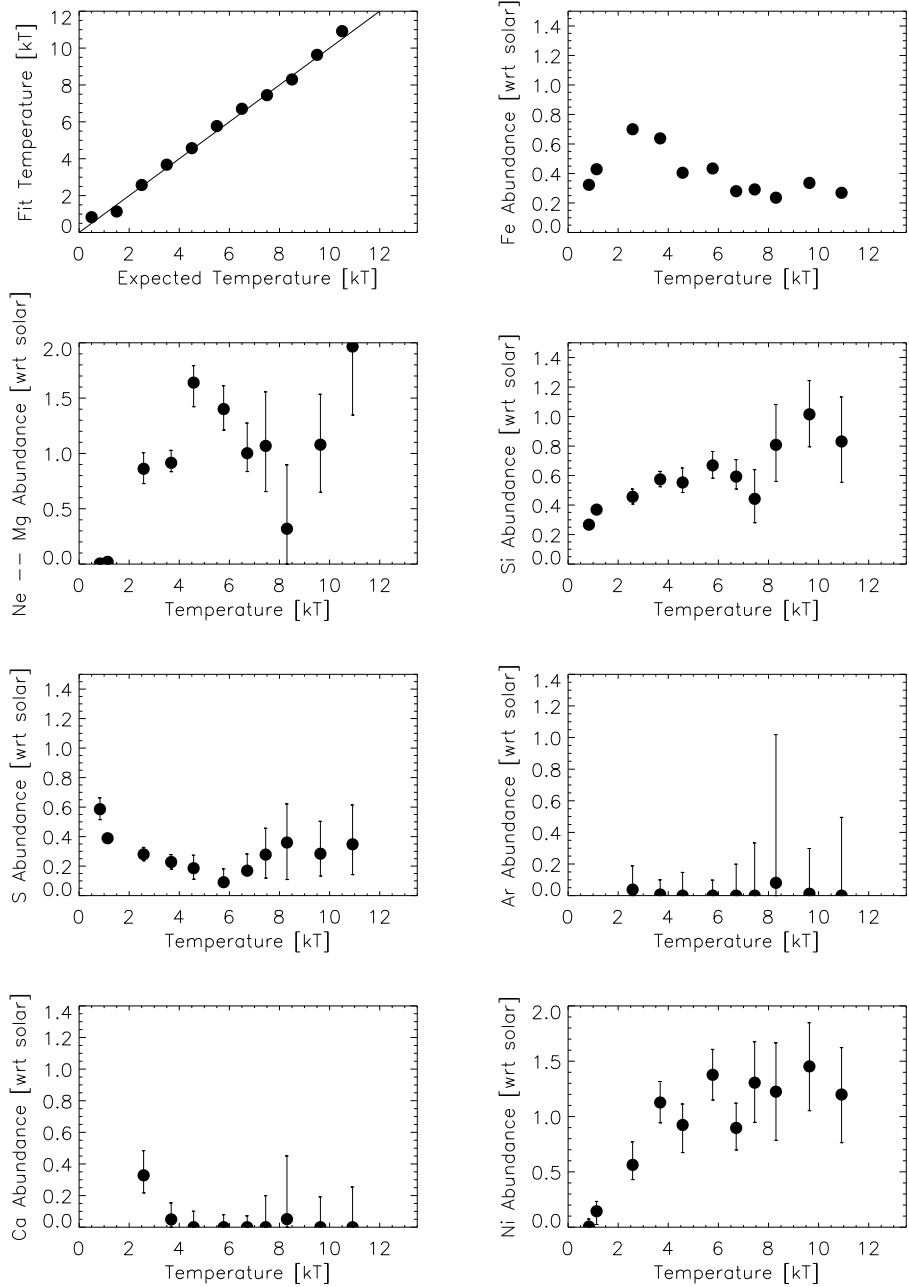}}}
\end{center}
\caption{The intermediate element abundances in galaxy clusters as a
function of temperature. Each point represents an average value for
several clusters with similar properties (the number of clusters in a
bin ranges from 13 -- 47).  The error bars are 90\% confidence
levels.  The error bars for iron are smaller than the plotted points.
}
\label{abuns}
\end{figure}

In Figure~\ref{starscompare} we compare our silicon and sulfur cluster
X-ray results to optical data from stars given in Timmes, Woosley, \&
Weaver (1995).  Our main results are given graphically in Figure~\ref{abuns}.
The most striking result is that the several $\alpha$ elements do
not track each other consistently and have very different overall
abundances.  Figure~\ref{abuns} shows that calcium and argon are not
detected, and have statistically significant upper limits, while
silicon and sulfur range from about 0.1 to 0.8 solar.  The abundances
for neon and magnesium are not considered well determined because the
main K-shell spectral lines for these elements fall in the middle of
the numerous lines from the L-shell of iron, confusing the analysis.
Also very perplexing is the fact that silicon rises with increasing
temperature, while sulfur falls.  All of these elements belong to the
elemental family formed by adding $\alpha$ particles to oxygen, and
are expected to have similar abundances as a result of their similar
formation processes.

\begin{figure}[t]
\resizebox{\textwidth}{!}{\rotatebox{90}{\includegraphics{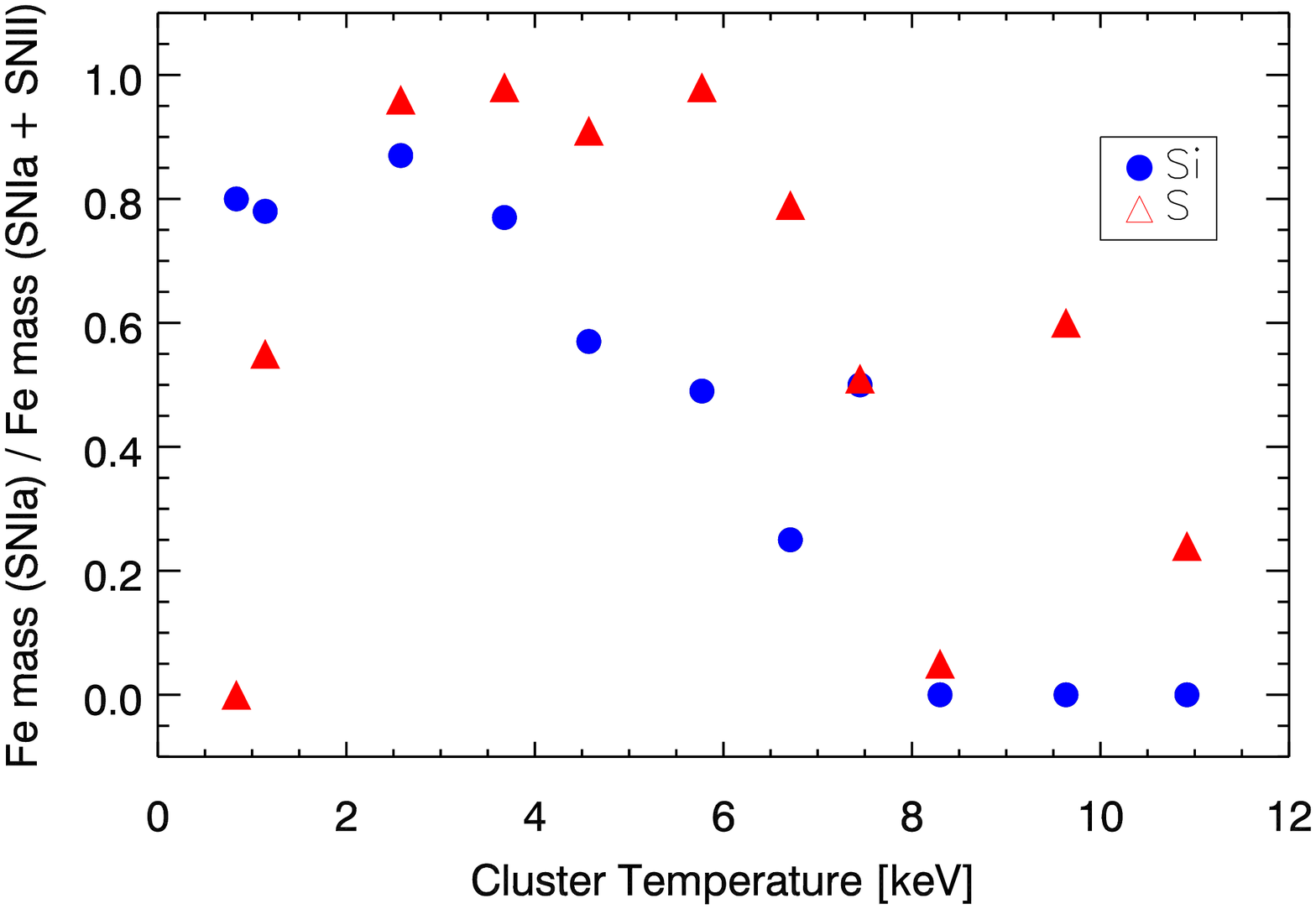}}}
\caption{The supernovae ratio in clusters.  This plot shows the SN
ratio based on the [S/Fe] and [Si/Fe] measurements and the W7 and
TNH-40 models for Type Ia and Type II supernovae.  A value of 1.0
indicates that all the iron mass in the ICM came from Type Ia, and a
value of 0.0 indicates that all the iron mass came from Type II.}
\label{sn_ratios}
\end{figure}

We have attempted to determine the supernovae (SN) fraction capable of
producing the observed metals in clusters by comparing the Si/Fe and
S/Fe ratios from theoretical yields for Type~Ia and Type~II SN given by 
the THN-40 and W7 models to our observed data.
Figure~\ref{sn_ratios} 
shows the SN ratio derived from our data for silicon and sulfur and
the theoretical yields.  Even at the same mass scale, the Type I to
Type II ratio determined by the elemental ratios depends on the choice
of elements used.  This suggests that the supernovae models need
refinement, that we need to consider mechanisms for distributing the
metals within the ICM, or both.  However, it can be said that low mass
clusters have a higher proportion of their iron from Type~Ia SN, and
that high mass clusters have most of their iron from Type~IIs.  This
supports the hypothesis that low mass groups do not have a large
enough gravitational potential to hold onto the energetic products
of Type~II SN.

\smallskip
\smallskip

Thanks go to Mike Loewenstein for useful discussions and access to his
collection of SN yields.


\end{document}